\documentclass{PoS}
\usepackage{braket}
\usepackage{amsmath}
\usepackage{amsbsy}
\usepackage{amsfonts}
\usepackage{amssymb}
\usepackage{graphics}
\usepackage{graphicx}
\usepackage{capt-of}

\title{Quantum-correlated measurements of $\boldsymbol{D^{0}\to K^{0}_{\rm S}\pi^{+}\pi^{-⁠}\pi^{0}}$ and consequences for the determination of $\gamma$}

\ShortTitle{Coherence of $D^{0}\to K^{0}_{\rm S}\pi^{+}\pi^{-⁠}\pi^{0}$ and consequences for the determination of $\gamma$}

\author{\speaker{Resmi PK}%
        \thanks{The CLEO collaboration is acknowledged for permitting the use of the data analysed in this study. This work is supported by UK-India Education and Research Initiative.}\\
       Indian Institute of Technology Madras\\
       E-mail: \email{resmipk@physics.iitm.ac.in}}

\author{Jim Libby\\
        Indian Institute of Technology Madras\\
        E-mail: \email{libby@iitm.ac.in}}
        
\author{Sneha Malde\\
        University of Oxford\\
        E-mail: \email{Sneha.Malde@physics.ox.ac.uk}}
\author{Guy Wilkinson\\
        University of Oxford\\
        E-mail: \email{Guy.Wilkinson@cern.ch}}        

\abstract{Quantum-correlated measurements of the decay $D^{0}~\to~K^{0}_{\rm S}\pi^{+}\pi^{-⁠}\pi^{0}$ are performed using a data sample corresponding to an integrated luminosity of 818 pb$^{-1}$ collected at the $\psi(3770)$ resonance by the CLEO-c detector. Preliminary results are presented for the $CP$-even fraction $F_{+}$ and the strong-phase differences of this decay. The value of $F_{+}$ is measured to be 0.246 $\pm$ 0.018. The strong-phase differences are measured in different regions of $K^{0}_{\rm S}\pi^{+}\pi^{-⁠}\pi^{0}$ phase space by binning around the intermediate resonances present. The potential sensitivity of the results for determining the CKM angle $\gamma$ from $B^{\pm}~\to~D(K^{0}_{\rm S}\pi^{+}\pi^{-}\pi^{0})K^{\pm}$ decay using the data collected by the Belle detector is also shown.}

\FullConference{ 9th International Workshop on the CKM Unitarity Triangle\\
         28  November - 3 December 2016\\
         Tata Institute for Fundamental Research (TIFR), Mumbai, India}

\begin{document}

\section{Introduction}

Among the three CKM \cite{CKM} angles $\gamma$ is measured least precisely. This is due to the small branching fraction of decays sensitive to $\gamma$. An improved measurement of $\gamma$ is essential for testing the standard model description of $CP$ violation. The decays $B^{\pm}\to DK^{\pm}$, where $D$ indicates a neutral charm meson reconstructed in a final state common to both $D^{0}$ and $\bar{D^{0}}$, provide $CP$-violating observables and they can be used for measuring $\gamma$ using data collected at detectors such as BaBar, Belle, LHCb or the future Belle II experiment. The additional inclusion of multibody  $D$ meson final states will reduce the statistical uncertainty on $\gamma$. However, multibody final states require knowledge of the strong-phase difference between the $D^0$ and $\bar{D}^{0}$ that varies over the phase space. The required strong-phase information can be obtained by studying quantum-correlated $D\bar{D}$ mesons produced in $e^{+}e^{-}$ collisions at an energy corresponding to $\psi(3770)$ at CLEO-c.

Here, we present preliminary results for the decay $D^{0}\to K_{\rm S}^{0}\pi^{+}\pi^{-}\pi^{0}$, which has a large branching fraction of 5.2$\%$ \cite{PDG}. This decay mode has not been used so far to determine  $\gamma$. The mode is potentially useful in a quasi-GLW \cite{GLW} analysis along with other $CP$ eigenstates if its $CP$-even fraction $F_{+}$ is known \cite{MNayak}. Further, this multibody self-conjugate decay occurs via many intermediate resonances, such as $K_{\rm S}^{0}\omega$ and $K^{*\pm}\rho^{\mp}$, hence if the strong-phase difference variation over the phase space is known, a GGSZ-style \cite{GGSZ, GGSZ2} analysis to determine $\gamma$ from this final state alone is possible.

\section{Quantum-correlated $D$ mesons}

The wave function for the decay of the vector meson $\psi$(3770) to a pair of $D$ mesons is antisymmetric as the two $D$ mesons are produced in a $P$-wave state. Integrating over the whole phase space, the double-tagged yield, where the decays of both the $D$ mesons are specified, for a signal (tag) decay $f$ ($g$) can be written in terms of the $CP$-even fraction $F_{+}^{f}$ $(F_{+}^{g})$ and the branching fractions $\mathcal{B}(f)$, $(\mathcal{B}(g))$ as
\begin{equation}
M(f|g) = \mathcal{N}\mathcal{B}(f)\mathcal{B}(g)\epsilon(f|g)\left[ 1 - (2F_{+}^{f} - 1) (2F_{+}^{g} - 1) \right],
\end{equation}
where $\mathcal{N}$ is the overall normalization factor and $\epsilon$ is the reconstruction efficiency. If $f$ or $g$ is a $CP$ eigenstate, then the value $(2F_{+} - 1)$ becomes the $CP$ eigenvalue $\lambda_{CP}$. So there is two-fold enhancement in the yield if $f$ and $g$ have opposite $CP$ eigenvalue and the yield becomes zero if $f$ and $g$ have the same $CP$ eigenvalue. Thus the rate of the decays of the two $D$ mesons are correlated to each other.

The single-tagged yield, where only one of the $D$ mesons is reconstructed without any constraints on the other, is given by
\begin{equation}
S(g) = \mathcal{N}\mathcal{B}(g) \epsilon(g).
\end{equation}
Assuming $\epsilon(f|g)=\epsilon(f)\epsilon(g)$, we write the ratios between the double-tagged and single-tagged yields, $N^{+}$ and $N^{-}$, when mode $g$ is a $CP$-odd ($\lambda_{CP}^{g}$ = $-1$) or $CP$-even ($\lambda_{CP}^{g}$ = 1), as
\begin{equation}
N^{\pm} = \frac{M(f|g)}{S(g)} = \mathcal{B}(f) \epsilon(f) \left[1\mp(2F_{+}^{f}-1)\right],
\end{equation}
which leads to the definition of $F_{+}^{f}$ in terms of $N^{+}$ and $N^{-}$:
\begin{equation}
 F_{+}^{f} \equiv \frac{N^{+}}{N^{+}+N^{-}}.
\end{equation}

In addition, we can also use some tag modes whose $CP$-even fraction $F_{+}^{g}$ is already known to determine $F_{+}^{f}$. For this, we define a quantity $N^{g}$ as the ratio of double-tagged and single-tagged yields as 
\begin{equation}
N^{g} = \mathcal{B}(f)\epsilon(f) \left[ 1-(2F_{+}^{f} - 1)(2F_{+}^{g} - 1) \right].
\end{equation}
This is used along with $N^{+}$ to extract $F_{+}^{f}$ as
\begin{equation}
F_{+}^{f} = \frac{N^{+}F_{+}^{g}}{N^{g} - N^{+} + 2N^{+}F_{+}^{g}}.\label{Eqn:PiPiPi0}
\end{equation}
The $g$ mode can also be self-conjugate modes like $K_{\rm S}^{0}\pi^{+}\pi^{-}$ or $K_{\rm L}^{0}\pi^{+}\pi^{-}$. The phase space of these multibody states can be divided into different bins. The $K_{\rm S,L}^{0}\pi^{+}\pi^{-}$ Dalitz plot is studied and binned according to the Equal $\delta_{D}$ scheme \cite{KsPiPi} based on the amplitude model reported in Ref. \cite{BaBar}. The double-tagged yield in each of these bins is 
\begin{equation}
M_{i}(K_{\rm S}^{0}\pi^{+}\pi^{-}\pi^{0} | K_{\rm S,L}^{0}\pi^{+}\pi^{-}) = h_{K_{\rm S,L}^{0}\pi^{+}\pi^{-}}(K_{i}^{K_{\rm S,L}^{0}\pi^{+}\pi^{-}}+K_{-i}^{K_{\rm S,L}^{0}\pi^{+}\pi^{-}}-2c_{i}\sqrt{K_{i}^{K_{\rm S,L}^{0}\pi^{+}\pi^{-}}K_{-i}^{K_{\rm S,L}^{0}\pi^{+}\pi^{-}}}(2F_{+}^{f}-1)),\label{Eqn:F+KhPiPi}
\end{equation}
where $K_{i}$ and $K_{-i}$ are the fraction of flavour-tagged $D^{0}$ and $\bar{D^{0}}$  decays in each bin,  $c_{i}$ is the cosine of the strong phase difference for $K_{\rm S,L}^{0}\pi^{+}\pi^{-}$, and $h_{K_{\rm S,L}^{0}\pi^{+}\pi^{-}}$ is the normalization factor. With these $F_{+}^{f}$ can be determined if the double-tagged yields in each of the $K_{\rm S,L}^{0}\pi^{+}\pi^{-}$ bins are measured.

To perform a GGSZ analysis with a self-conjugate multibody final state $f$, the amplitude-weighted averages of $\cos\Delta\delta_{D}$ and $\sin\Delta\delta_{D}$ over regions of phase space \cite{GGSZ, GGSZ2}, referred to as $c_{i}$ and $s_{i}$, respectively are required. Here $\Delta\delta_D$ is the strong-phase difference between $CP$ conjugate points in the phase space. The values of $c_i$ and $s_i$ are obtained by tagging with $CP$ and quasi-$CP$ eigenstates and other self-conjugate modes. For $CP$ eigenstate tag modes, the double-tagged yield is given by
\begin{equation}
M_{i}^{\pm} = h_{CP}\left[ K_{i}+\bar{K_{i}} \mp 2 \sqrt{K_{i} \bar{K_{i}}} c_{i} \right],\label{Eqn:CP}
\end{equation}
where $h_{CP}$ is the normalization constant. If the tag is a quasi-$CP$ eigenstate of known $F_{+}$, the $c_{i}$ sensitive term is scaled by ($2F_{+} - 1$) rather than 1. For the self-conjugate tag mode $K_{\rm S}^{0}\pi^{+}\pi^{-}$ \cite{KsPiPi:EPJC1, KsPiPi:EPJC2}, the double-tagged yield is 
\begin{footnotesize}
\begin{equation}
M_{i\pm j}^{K_{\rm S}^{0}\pi^{+}\pi^{-}} = h_{K_{\rm S}^{0}\pi^{+}\pi^{-}} \left[ K_{i} K_{\mp j}^{K_{\rm S}^{0}\pi^{+}\pi^{-}} + \bar{K_{i}} K_{\pm j}^{K_{\rm S}^{0}\pi^{+}\pi^{-}} - 2 \sqrt{K_{i} K_{\pm j}^{K_{\rm S}^{0}\pi^{+}\pi^{-}}\bar{K_{i}}K_{\mp j}^{K_{\rm S}^{0}\pi^{+}\pi^{-}}} ( c_{i}c_{j}^{K_{\rm S}^{0}\pi^{+}\pi^{-}} + s_{i}s_{j}^{K_{\rm S}^{0}\pi^{+}\pi^{-}})\right],
\end{equation}
\end{footnotesize}
 and for a  $K_{\rm L}^{0}\pi^{+}\pi^{-}$ tag, the double-tagged yield is 
\begin{footnotesize}
\begin{equation}
M_{i\pm j}^{K_{\rm L}^{0}\pi^{+}\pi^{-}} = h_{K_{\rm L}^{0}\pi^{+}\pi^{-}} \left[ K_{i} K_{\mp j}^{K_{\rm L}^{0}\pi^{+}\pi^{-}} + \bar{K_{i}} K_{\pm j}^{K_{\rm L}^{0}\pi^{+}\pi^{-}} + 2 \sqrt{K_{i} K_{\pm j}^{K_{\rm L}^{0}\pi^{+}\pi^{-}}\bar{K_{i}}K_{\mp j}^{K_{\rm L}^{0}\pi^{+}\pi^{-}}} ( c_{i}c_{j}^{K_{\rm L}^{0}\pi^{+}\pi^{-}} + s_{i}s_{j}^{K_{\rm L}^{0}\pi^{+}\pi^{-}})\right],
\end{equation}
\end{footnotesize}
where $h_{K_{\rm S,L}^{0}\pi^{+}\pi^{-}}$ are the normalization constants.
If both tag and signal states are the same, then 
\begin{equation}
M_{ij} = h_{f} \left[ K_{i}\bar{K_{j}} + \bar{K_{i}} K_{j} - 2 \sqrt{K_{i} \bar{K_{j}} \bar{K_{i}} K_{j}} (c_{i}c_{j} + s_{i}s_{j}) \right ],\label{Eqn:DT}
\end{equation}
where $h_{f}$ is the normalization constant.

\section{Measurement of $F_{+}$ in $D^{0}\to K_{\rm S}^{0}\pi^{+}\pi^{-}\pi^{0}$ decays }

A data sample corresponding to an integrated luminosity of 818 pb$^{-1}$, collected by the CLEO-c detector at the interaction point of CESR $e^{+}e^{-}$ collider, consisting of $D\bar{D}$ pairs coming from the $\psi(3770)$ resonance is used in this analysis. The $D\bar{D}$ final state is reconstructed for the signal state $K_{\rm S}^{0}\pi^{+}\pi^{-}\pi^{0}$ along with the tag modes given in Table~\ref{Table:Tags}. All tracks and showers associated with both the $D$ mesons are reconstructed; the selection criteria for the tag modes are identical to those presented in Ref. \cite{MNayak}. Modes involving $K_{\rm L}^{0}$ or $\nu$ are reconstructed partially using a missing-mass squared technique \cite{MissMass}.

\begin{table} [t] 
\centering  
 \begin{tabular} {c c  }
\hline 
Type & Modes \\[0.5ex]
\hline
\hline
 $CP$-even & $K^{+}K^{-}$, $\pi^{+}\pi^{-}$, $K_{\rm S}^{0}\pi^{0}\pi^{0}$, $K_{\rm L}^{0}\omega$, $K_{\rm L}^{0}\pi^{0}$ \\[0.5ex]
 $CP$-odd & $K_{\rm S}^{0}\pi^{0}$, $K_{\rm S}^{0}\eta$, $K_{\rm S}^{0}\eta'$ \\[0.5 ex]
 Mixed $CP$ & $\pi^{+}\pi^{-}\pi^{0}$, $K_{\rm S}^{0}\pi^{+}\pi^{-}$, $K_{\rm L}^{0}\pi^{+}\pi^{-}$ \\[0.5 ex]
 Flavour & $K^{\pm}e^{\mp}\nu_{\rm e}$ \\[0.5ex]
\hline
\end{tabular}  
\caption{Different tag modes used in the analysis.}\label{Table:Tags}
\end{table} 

\begin{figure}[t]
\centering
\begin{tabular}{cc}
\includegraphics[width=7cm]{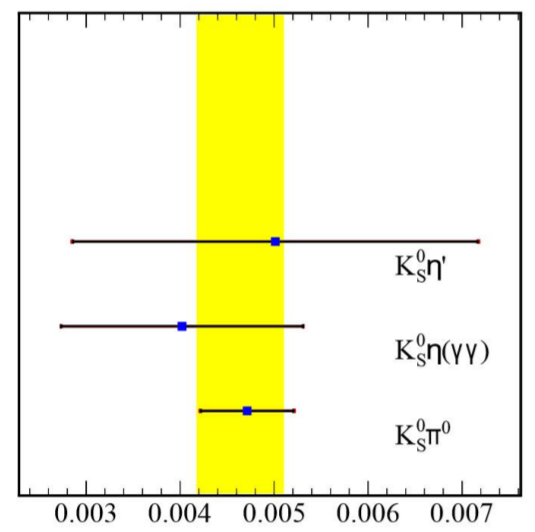}&
\includegraphics[width=7cm]{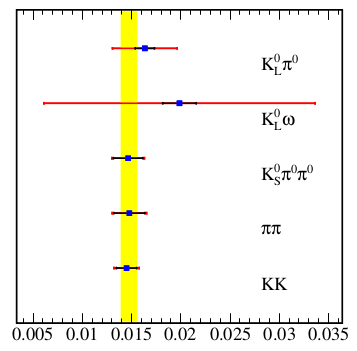}\\
\end{tabular}
\caption{ $N^{+}$ values for the $CP$-odd modes (left) and $N^{-}$ values for the $CP$-even modes (right). The yellow region shows the average value. Horizontal black lines show the statistical uncertainty and red lines the total uncertainty.}\label{Fig:N}
\end{figure}

\begin{figure}[ht!]
\begin{center}
\begin{tabular}{cc}
\includegraphics[width=7.5cm]{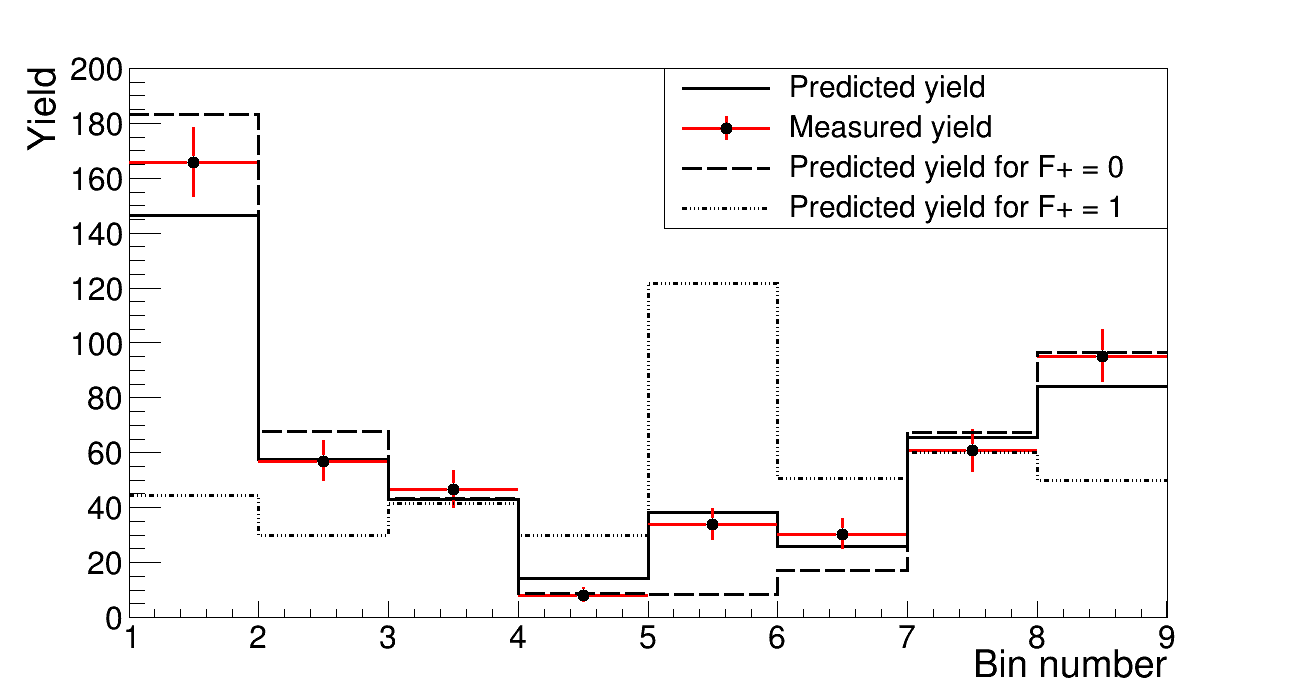}&
\includegraphics[width=7.5cm]{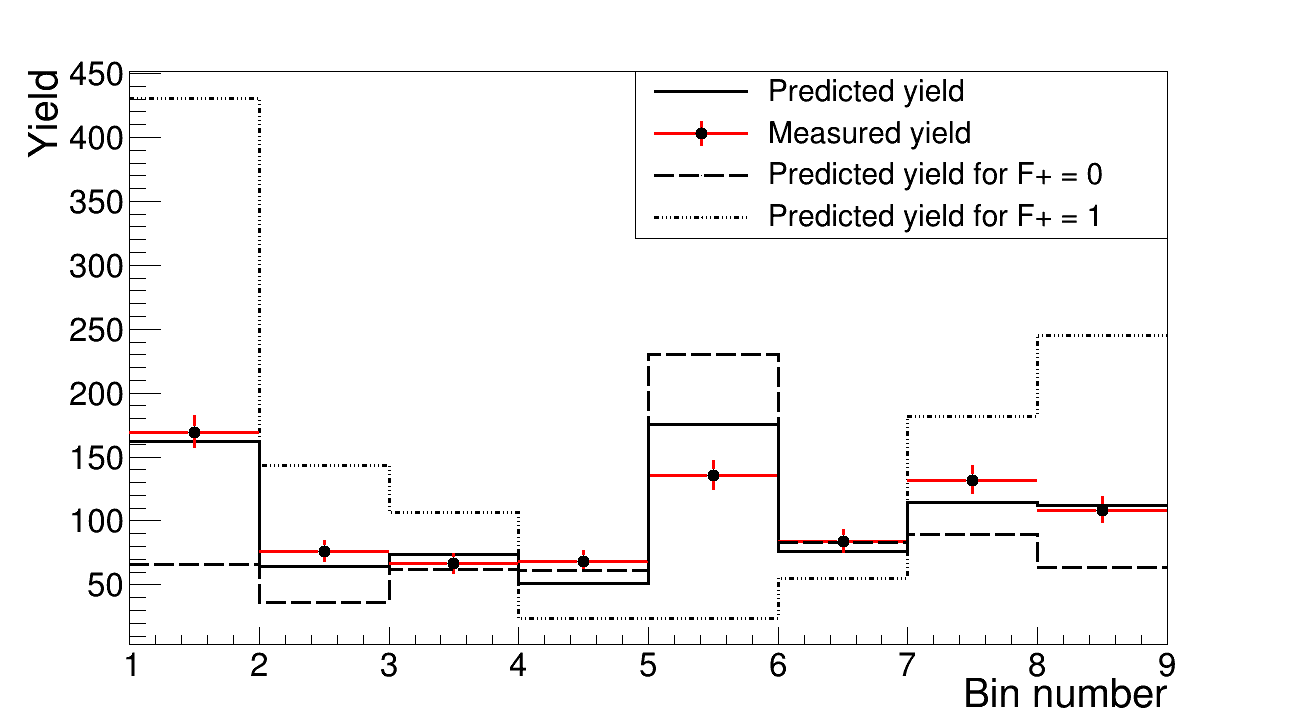}\\
\end{tabular}
\end{center}
\caption{The predicted and measured yields for $K_{\rm S}^{0}\pi^{+}\pi^{-}$ (left) and  $K_{\rm L}^{0}\pi^{+}\pi^{-}$ (right) in each bin obtained from the combined fit of both the modes. The histogram shows the predicted values, points show the measured values, dashed line corresponds to $F_{+}$ = 0 and the dotted line shows $F_{+}$ = 1.}\label{Fig:KhPiPi}
\end{figure}

With the double-tagged yields measured and single-tagged yields taken from Ref. \cite{4pi}, we calculate $N^{+}$ and $N^{-}$ from the $CP$-odd and $CP$-even modes respectively. They are shown in Fig.~\ref{Fig:N}. With the quasi-$CP$ mode $\pi^{+}\pi^{-}\pi^{0}$, we calculate $F_{+}$ using Eqn.~\ref{Eqn:PiPiPi0} with input value $F_{+}^{\pi^{+}\pi^{-}\pi^{0}}$ = 0.973~$\pm$~0.017 \cite{4pi}. The value of $F_{+}$  obtained with $CP$ and quasi-$CP$ modes is 0.244~$\pm$~0.021. This suggests that the mode $K_{\rm S}^{0}\pi^{+}\pi^{-}\pi^{0}$ is significantly $CP$-odd. Using $K_{\rm S,L}^{0}\pi^{+}\pi^{-}$ modes, $F_{+}$ is calculated with Eqn.~\ref{Eqn:F+KhPiPi}. The values of $K_{i}$, $K_{-i}$, $c_{i}$, and $s_{i}$ for $K_{\rm S,L}^{0}\pi^{+}\pi^{-}$ are taken from Ref.~\cite{KsPiPi}. The values of predicted and measured double-tagged yields in each of the $K_{\rm S,L}^{0}\pi^{+}\pi^{-}$ bins are shown in Fig.~\ref{Fig:KhPiPi}; from these data $F_{+}$ is determined to be 0.265~$\pm$~0.029 in this calculation. With all the three above mentioned methods, the average $F_{+}$ is 0.246~$\pm$~0.018. The uncertainty includes statistical as well as systematic contributions.

\section{Determination of $c_{i}$ and $s_{i}$}
The five-dimensional phase space of $D^{0}\to K_{\rm S}^{0}\pi^{+}\pi^{-}\pi^{0}$ is studied to extract $c_{i}$ and $s_{i}$ values. There is no trivial symmetry in the phase space to define the bins and hence the bins are constructed around the resonances present.  The lack of an amplitude model for this channel makes a proper optimization difficult. An exclusive eight-bin scheme is followed around the resonances such as $\omega$, $K^{*}$ and $\rho$. The kinematic regions of the bins are given in Table~\ref{Table:Bin} along with the fraction of flavour-tagged $D^{0}$ and $\bar{D^{0}}$ decays in each of them. These values are determined from semileptonic flavour tag $K^{\pm}e^{\mp}\nu_{\rm e}$.
\begin{table}[ht!] 
\centering
\begin{tabular} {c c c c}
\hline \\ [0.1ex]
 Bin number & Specification & $K_{i}$ & $\bar{K_{i}}$  \\[0.5ex]
\hline
\hline
 1 & m($\pi^{+}\pi^{-}\pi^{0}$) $\approx$ m($\omega$) & $0.222\pm0.019$ & $0.176\pm0.017$\\[0.5ex]
 2 & m($K_{\rm S}^{0}\pi^{-}$) $\approx$ m($K^{*-}$) $\&$ m($\pi^{+}\pi^{0}$) $\approx$ m($\rho^{+}$) & $0.394\pm0.022$ & $0.190\pm0.017$ \\[0.5ex]
 3 & m($K_{\rm S}^{0}\pi^{+}$) $\approx$ m($K^{*+}$) $\&$ m($\pi^{-}\pi^{0}$) $\approx$ m($\rho^{-}$) & $0.087\pm0.013$ & $0.316\pm0.021$ \\[0.5ex]
 4 & m($K_{\rm S}^{0}\pi^{-}$) $\approx$ m($K^{*-}$) & $0.076\pm0.012$ & $0.046\pm0.009$ \\[0.5ex]
 5 & m($K_{\rm S}^{0}\pi^{+}$) $\approx$ m($K^{*+}$) & $0.057\pm0.010$ & $0.065\pm0.011$ \\[0.5ex]
 6 & m($K_{\rm S}^{0}\pi^{0}$) $\approx$ m($K^{*0}$) & $0.059\pm0.011$ & $0.092\pm0.013$ \\[0.5ex]
 7 & m($\pi^{+}\pi^{0}$)  $\approx$ m($\rho^{+}$) & $0.045\pm0.009$ & $0.045\pm0.009$\\[0.5ex]
 8 & Remainder & $0.061\pm0.011$ & $0.070\pm0.011$\\[0.5ex]
\hline
\end{tabular} 
\caption{The specifications for the eight exclusive bins of $D^{0}\to K_{\rm S}^{0}\pi^{+}\pi^{-}\pi^{0}$ phase space along with the fraction of $D^{0}$ and $\bar{D^{0}}$ events in each of them. }\label{Table:Bin}
\end{table}

The yields for $CP$, quasi-$CP$ and self-conjugate modes in each of the bins are measured and the $c_{i}$ and $s_{i}$ values are found out using Eqn. \ref{Eqn:CP}-\ref{Eqn:DT}. The migration of events from one bin to another due to the narrowness of each bin is considered in the fit. The preliminary results are summarized in Table~\ref{Table:cisi} and Fig.~\ref{Fig:cisi}. The uncertainties mentioned are statistical only.

\begin{table}[ht]
\begin{minipage}[b]{0.5\linewidth}
\centering
\begin{tabular} {c  c c }
\hline 
Bin & $c_{i}$ & $s_{i}$\\[0.5ex]
\hline
\hline
 1&         $-$1.12 $\pm$ 0.12& ~~0.12 $\pm$ 0.17\\[0.5ex]
 2&     $-$0.29 $\pm$ 0.07 & ~~0.11 $\pm$ 0.13 \\[0.5ex]
 3&          $-$0.41 $\pm$ 0.09 & $-$0.08 $\pm$ 0.18 \\[0.5ex]
 4&         $-$0.84 $\pm$ 0.12& $-$0.73 $\pm$ 0.34 \\[0.5ex]
 5&           $-$0.54 $\pm$ 0.13 & ~~0.65 $\pm$ 0.13 \\[0.5ex]
 6&          $-$0.22 $\pm$ 0.12& ~~1.37 $\pm$ 0.22 \\[0.5ex]
 7&           $-$0.90 $\pm$ 0.16&  $-$0.12 $\pm$ 0.40 \\[0.5ex]
 8&           $-$0.70 $\pm$ 0.14 &  $-$0.03 $\pm$ 0.44 \\[0.5ex]
\hline
\end{tabular}
\caption{The preliminary results for $c_{i}$ and $s_{i}$ values obtained from the fit.}\label{Table:cisi}
\end{minipage}\hfill
\begin{minipage}[b]{0.5\linewidth}
\centering
\includegraphics[width=6.3cm, height=6cm]{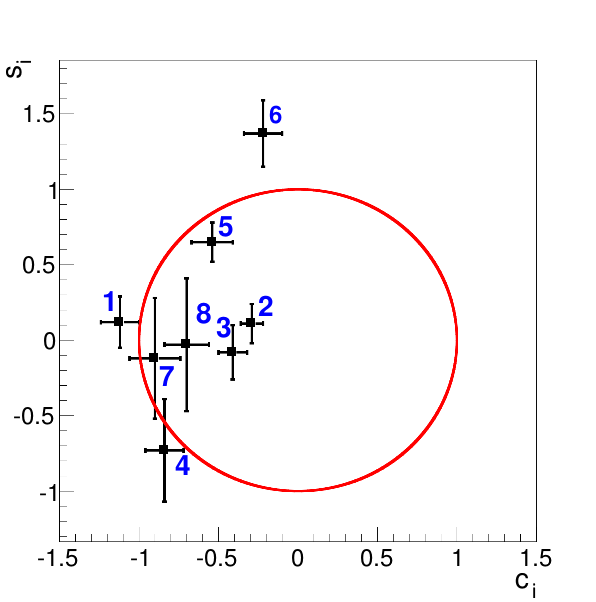}
\captionof{figure}{$c_{i}$ and $s_{i}$ values in each bin.}\label{Fig:cisi}
\end{minipage}
\end{table}

\section{Estimation of $\gamma$ sensitivity with $B^{\pm}\to D(K^{0}_{\rm S}\pi^{+}\pi^{-}\pi^{0})K^{\pm}$ }

\begin{figure}[ht!]
\centering
\includegraphics[width=6cm, height=5cm]{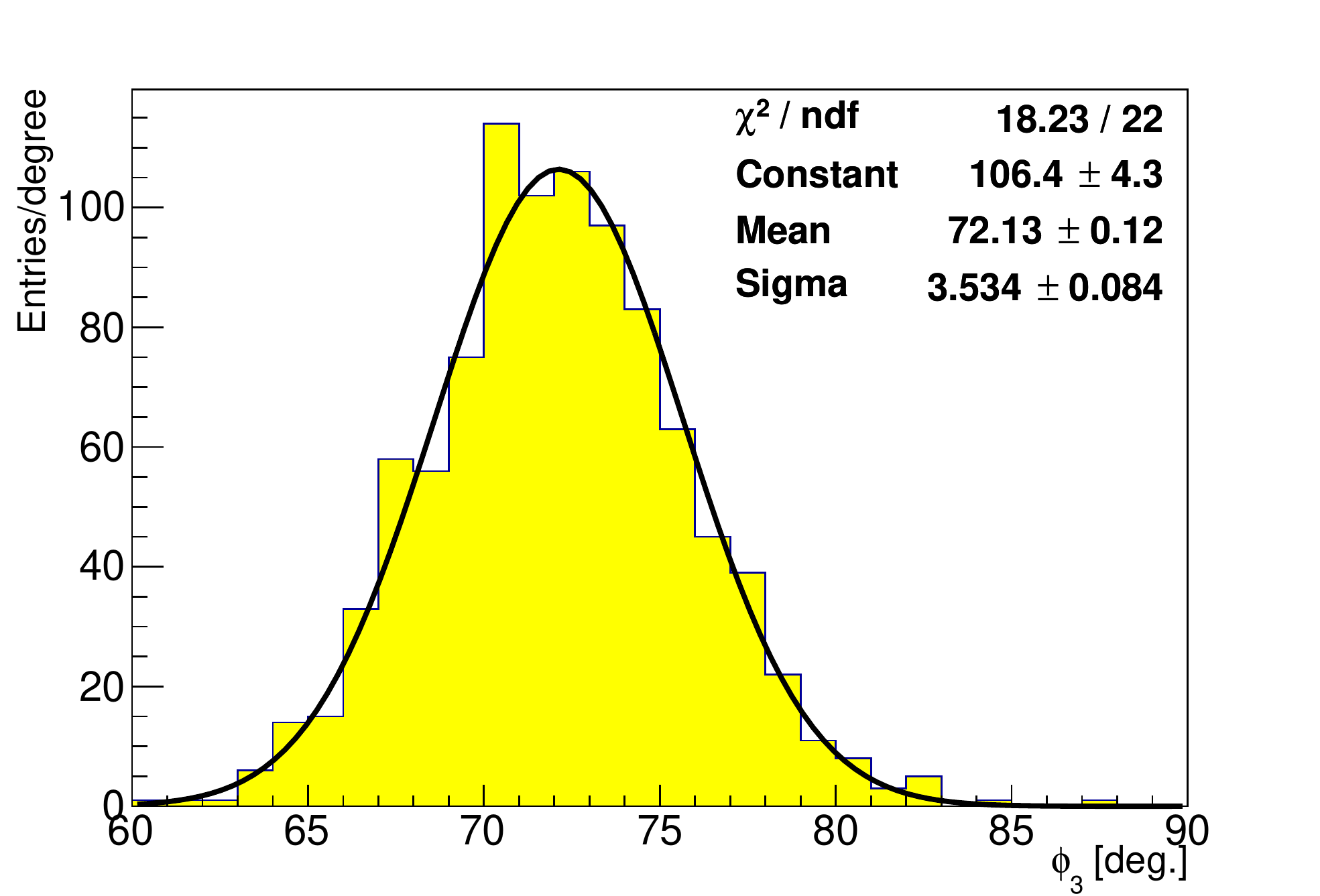}
\vspace{-0.1 in}
\caption{$\gamma$ sensitivity with 50 $\mathrm{ab}^{-1}$ Belle II sample. }\label{Fig:sensitivity}
\end{figure}

We estimate the sensitivity of $\gamma$ with the preliminary results of $c_{i}$ and $s_{i}$ values described in the previous section, in a GGSZ framework with $B^{\pm}\to D(K^{0}_{\rm S}\pi^{+}\pi^{-}\pi^{0})K^{\pm}$ decays from Belle ($\approx$ 1~$\mathrm{ab}^{-1}$). We run 1000 pseudo experiments with $c_i$, $s_i$, $K_i$, and $\bar{K_i}$ values as inputs with each experiment consisting of $\approx$ 1200 events. The sample sizes are determined from the Belle sample of $B^{\pm}\to D(K^{0}_{\rm S}\pi^{+}\pi^{-})K^{\pm}$ \cite{Belle-GGSZ}. Here we assume that increase in branching fraction for $K^{0}_{\rm S}\pi^{+}\pi^{-}\pi^{0}$ compared to $K^{0}_{\rm S}\pi^{+}\pi^{-}$ is compensated by loss of efficiency due to a $\pi^{0}$ in final state. The estimated uncertainty on $\gamma$ is $\sigma_{\gamma}=25^{\circ}$. The projection of this to a 50~$\mathrm{ab}^{-1}$ sample of Belle II gives $\sigma_{\gamma}=3.5^{\circ}$ (see Fig.~\ref{Fig:sensitivity}). 

\section{Conclusions}

The studies of $D$ meson final states opens up additional ways of measuring $\gamma$. The decay $D^{0} \rightarrow K_{\rm S}^{0}\pi^{+}\pi^{-}\pi^{0}$ can serve as an additional mode in quasi-GLW methods with the $CP$-even fraction $F_{+}$ measured to be 0.246~$\pm$~0.018, reducing the statistical uncertainty on $\gamma$. In addition, the measurement of strong phase differences of this mode in eight different phase space regions, allows a model-independent GGSZ estimation of $\gamma$ from this mode alone. It is estimated that a single-mode uncertainty on $\gamma$ of $\sigma_{\gamma}=3.5^{\circ}$ is achievable with a 50~$\mathrm{ab}^{-1}$ sample of data at Belle II. This could be improved with optimized $c_{i}$ and $s_{i}$ values provided a proper amplitude model is available and a finer binning using a larger sample of quantum correlated data from the BESIII experiment.

\acknowledgments
We acknowledge the erstwhile CLEO collaboration members for the privilege of using the data for the analysis presented. We would like to thank UK-India Education and Research Initiative, IIT Madras, and TIFR Mumbai for being able to successfully attend this conference.

\end{document}